# Reverse Chromatic Aberration and its Numerical Optimization in a Metamaterial Lens

**William J. Capecchi\*, Nader Behdad, and Francesco A. Volpe**
*University of Wisconsin-Madison, Madison, WI 53706 USA*
*\*capecchi@wisc.edu*

**Abstract:** In planar metamaterial lenses, the focal point moves with the frequency. Here it is shown numerically that this movement can be controlled by properly engineering the dimensions of the metamaterial-based phase shifters that constitute the lens. In particular, such lenses can be designed to exhibit unusual chromatic aberration with the focal length increasing, rather than decreasing, with the frequency. It is proposed that such an artificial "reverse" chromatic aberration may optimize the transverse resolution of millimeter wave diagnostics of plasmas and be useful in compensating for the natural "ordinary" chromatic aberration of other components in an optical system. More generally, optimized chromatic aberration will allow to simultaneously focus on several objects located at different distances and emitting or reflecting at different frequencies.

---

---

## 1. Introduction

Many optical systems suffer from chromatic aberration (CA). A typical lens focuses higher frequencies $f$ closer to the lens and lower frequencies further away, due to the fact that the refractive index typically increases with frequency. A lens exhibiting "reverse" CA (RCA), meaning that it focuses higher $f$ at longer focal lengths $l$, could be used to correct for CA in these systems.

There are also cases in which finite amounts of CA or RCA are desirable but need to be optimized. Some of these cases can be found in millimeter wave diagnostics [1-4] of plasmas [5]: electron cyclotron emission depends on the local magnetic field; as a result, different *f* are emitted at different locations in a non-uniformly magnetized plasma. Similarly, externally injected waves of different frequency *f* are reflected at different locations in a plasma of non-uniform density. These considerations translate in *f*-dependent focusing requirements: for the highest resolution, the focal length *l* (of the order of 0.5-1.5m) needs to *increase* with *f* (typical central frequency: 4-140GHz) by up to 2cm/GHz over ranges $\Delta f$ of several GHz or tens of GHz.

Finally there are cases, for instance in photography and surveillance, in which objects located at different distances and emitting or reflecting at different frequencies need to be imaged on the same focal plane.

RCA was recently observed in a metamaterial lens [6] developed for phased array antenna applications at 8-12GHz. RCA is an undesired property for that application. In this work, we examine metamaterial lenses of the type reported in [6] and numerically optimize them for plasma diagnostics in the same range of frequencies. The optimization method, however, can be easily extended to the other applications mentioned above, as well as to different frequencies, including visible, where similar geometries are utilized, although not optimized for RCA.

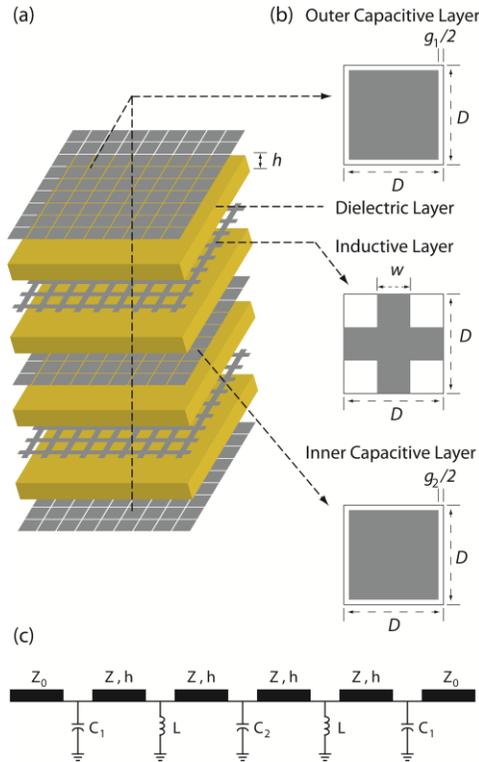

Fig. 1. Topology of a third-order bandpass MEFSS. (a) Exploded view of overall lens design. (b) Top views of a unit cell capacitive and inductive layers. (c) Equivalent circuit model valid for normal incidence.

Each lens consists of several spatial phase shifters, or pixels, that populate its aperture. Each spatial phase shifter is the unit cell of a miniaturized-element frequency selective surface (MEFSS) of the type reported in [7]. MEFSSs are composed of periodic arrangements of sub-wavelength capacitive patches and inductive wire grids. An MEFSS composed of $N$ capacitive layers alternated by $N-1$ inductive layers acts as an $N^{th}$-order coupled-resonator bandpass filter [7]. Fig. 1a-b shows an exploded view of an MEFSS with a $3^{rd}$-order bandpass response. The phase delay provided by an MEFSS is controllable by varying the physical and geometrical parameters of its constituting unit cell (e.g., dimensions of the capacitive and inductive layers, the thickness and material of the dielectric layers, etc.). In the design considered in this paper, each cell has a width $D$=6.1mm which is much smaller than the wavelength ($\lambda$=30mm at $f$=10GHz). Fig. 2 illustrates the benefit of this feature. For a typical lens, a smooth variation in phase delay can be achieved by light rays at different distances from the optical axis experiencing different path lengths through the lens (Fig. 2a). In constructing a lens with discrete electrically *large* unit cells (e.g. microwave lenses based on arrays of receiving antennas connected to arrays of transmitting antennas using transmission lines with variable lengths) the phase delay does not change smoothly with the distance from the optical axis (Fig. 2b). However, the continuous nature of a traditional lens is regained in the limit that cells much *smaller* than the wavelength can be constructed and assembled (Fig. 2c).

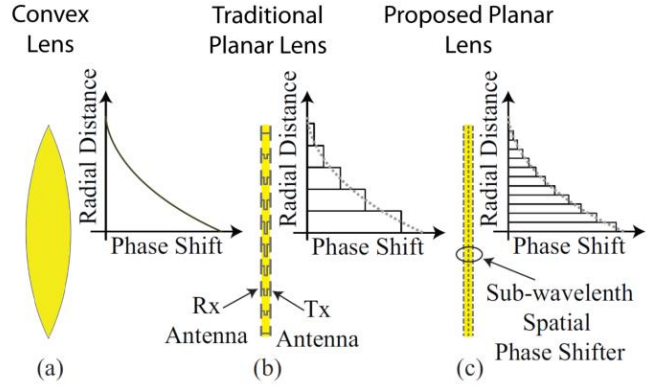

Fig. 2. The phase delay profile of a typical glass lens is a smooth function of radial distance (a). This is lost in discretized lens systems (b) but regained in the limit of small aperture features (c).

A better understanding of the operation of this structure can be gained by considering the equivalent circuit in Fig. 1c, which is valid for normal incidence. The patches in the first, third, and fifth metallic layers are modeled with parallel capacitors and the wire grid layers are modeled with parallel inductors. The dielectric substrates separating these layers are modeled with short transmission lines of length $h$ equal to the thickness of the dielectric in Fig. 1a and characteristic impedance $Z = Z_0/\sqrt{\epsilon_r}$, where $\epsilon_r$ is the dielectric constant of the substrate and $Z_0 = 377\Omega$ is the free space impedance. Free space on each side of the cell is modeled with semi-infinite transmission lines with characteristic impedances $Z_0$.

The effective capacitance of a 2-D periodic array of square metallic patches separated by gaps $g_1$ as in Fig. 1 is [7]:

$$C_1 = \epsilon_0 \epsilon_{eff} \frac{2D}{\pi} \ln\left[\frac{1}{\sin\left(\frac{\pi g_1}{2D}\right)}\right] \quad (1)$$

where $\epsilon_{eff}$ is the effective permittivity of the medium in which the capacitive patches are immersed. A similar expression relates the capacitance $C_2$ of the inner capacitive layer to the spacing $g_2$ between adjacent capacitive patches in that layer.

The effective inductance of a periodic wire mesh as pictured in Fig. 1a-b is described by [7]:

$$L = \mu_0 \mu_{eff} \frac{D}{2\pi} \ln\left[\frac{1}{\sin\left(\frac{\pi w}{2D}\right)}\right] \qquad (2)$$

Here $\mu_0$ is the free space permeability, $\mu_{eff}$ is the effective permeability of the medium, and $w$ is the strip width (Fig. 1b). Equations (1)-(2) are valid only when the structure is placed in a homogenous medium away from metallic objects and when the parameters $w$, $g_1$ and $g_2$ are not varied over the array. The similarity between this circuit model and a 3$^{rd}$ order coupled resonator frequency bandpass filter can be seen by replacing the transmission line sections with their equivalent circuit models, composed of a series inductor and shunt capacitor. A discussion of this can be found in [7].

A wave passing through the metamaterial lens in Fig. 1a experiences a frequency-dependent attenuation and phase shift. These can be calculated from the transfer function describing the equivalent circuit in Fig. 1c:

$$T = \frac{2}{A + \frac{B}{Z_0} + CZ_0 + D} \qquad (3)$$

where *A*, *B*, *C* and *D* are given by

$$\begin{bmatrix} A & B \\ C & D \end{bmatrix} = \begin{bmatrix} 1 & 0 \\ j\omega C_1 & 1 \end{bmatrix} \begin{bmatrix} \cos\beta h & jZ\sin\beta h \\ \frac{j}{Z}\sin\beta h & \cos\beta h \end{bmatrix} \begin{bmatrix} 1 & 0 \\ \frac{1}{j\omega L} & 1 \end{bmatrix} \begin{bmatrix} \cos\beta h & jZ\sin\beta h \\ \frac{j}{Z}\sin\beta h & \cos\beta h \end{bmatrix} \begin{bmatrix} 1 & 0 \\ j\omega C_2 & 1 \end{bmatrix} \times$$
$$\begin{bmatrix} \cos\beta h & jZ\sin\beta h \\ \frac{j}{Z}\sin\beta h & \cos\beta h \end{bmatrix} \begin{bmatrix} 1 & 0 \\ \frac{1}{j\omega L} & 1 \end{bmatrix} \begin{bmatrix} \cos\beta h & jZ\sin\beta h \\ \frac{j}{Z}\sin\beta h & \cos\beta h \end{bmatrix} \begin{bmatrix} 1 & 0 \\ j\omega C_1 & 1 \end{bmatrix} \qquad (4)$$

Here $Z$ and $\beta = \frac{\omega/c}{\sqrt{\varepsilon_r}}$ are respectively the characteristic impedance and propagation constant of the transmission line. The substitution of Eqs.1-2 in Eqs.3-4 and extraction of phase $\phi$ from Eq.3 yields a conceptually straightforward, but algebraically complicated relationship between $\phi$, $f$, $w$, $g_1$ and $g_2$, that suggests that the frequency dependence of $\phi$ can be altered by varying the geometrical parameters $w$, $g_1$ and $g_2$. That complicated expression, however, is approximate, in the measure in which Fig.1c is only an approximate description of the array in Fig.1a.

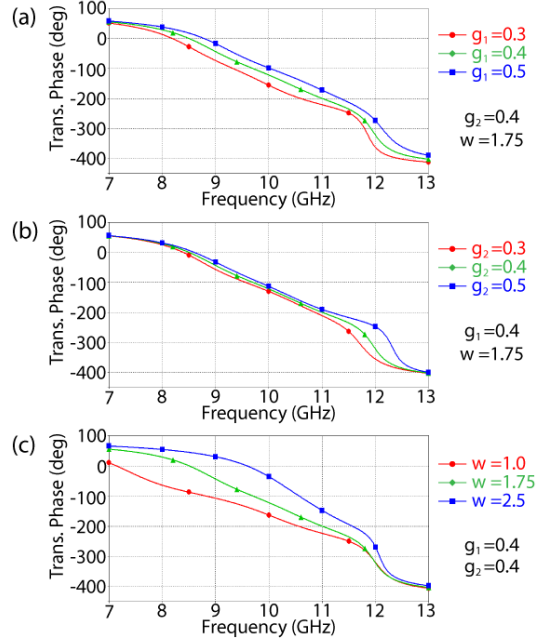

Fig. 3. Parameter sweeps highlighting the behavior of a 3$^{rd}$ order MEFSS when varying (a) $g_1$ (b) $g_2$ or (c) $w$, all the rest remaining fixed.

Instead, for more realistic estimates (for example to take into account the finite number of cells), the device in Fig. 1a was modeled numerically for various choices of $g_1$, $g_2$, and $w$, and the corresponding transmission and phase shift were calculated using full-wave EM simulations in CST Microwave Studio. Fig.3 illustrates how the function $\phi(f)$ varies with $g_1$ (Fig.3a), $g_2$ (Fig.3b) and $w$(Fig.3c), all the rest remaining fixed. Fig.4 illustrates how $\phi$ at a fixed $f$ varies with $g_1$, $g_2$, and $w$.

This discussion and numerical optimization applies to a single, uniform array. Then, similar to Fig. 2c, we combine seven concentric "zones" about the optical axis. Each zone is separately optimized to yield a certain phase shift as a function of the frequency, $\phi(f)$.

Note in Fig.4 that in general there are many (in fact, infinite) sets of $g_1$, $g_2$, $w$ solutions yielding the desired phase-shift $\phi$ for a certain frequency and zone. However, note also that, for each zone, different phases are required for each frequency.

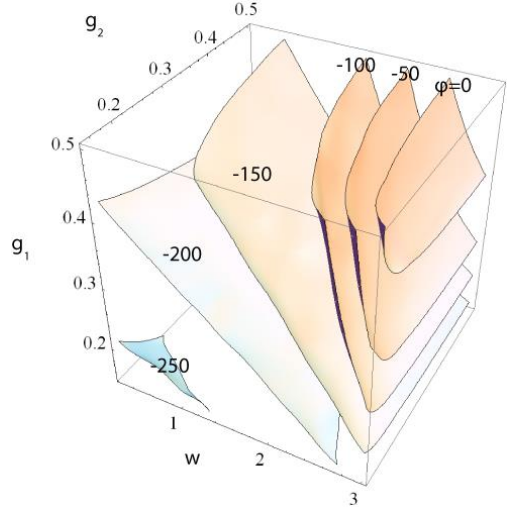

Fig 4. 3D iso-phase contour plot over $g_1$, $g_2$, and $w$ for $f=10$GHz. Note that for every desired $\varphi$, parameter solution $g_1$, $g_2$, $w$ is not unique for a given $f$. However, constraining $\varphi$ at multiple $f$ reduces the choices of $g_1$, $g_2$, and $w$.

## 2. Numerical Optimization

The starting point for our RCA optimization was an existing experimentally and numerically well-characterized 3$^{rd}$ order MEFSS design with a bandpass of 9-11GHz [6-8].

For proof of principle, we initially set the goal of modifying that design in such a way that it had a focal length $l=30$cm at $f=8$GHz and $l=126$cm at $f=12$GHz. These focal lengths were chosen since similar focal lengths will be needed in the DIII-D [9] tokamak [5], but at 10 times higher frequencies ($f=80$-130GHz) [10]. The relative variation of focal length considered here is the same. The miniaturization and other issues associated with the higher frequencies are left as future work. The requirement on the dependence of $l$ on $f$ translates into requirements on the $\phi(f)$ dependencies in the various "zones" [6].

The order of an MEFSS puts a constraint on the aperture size. For a 3$^{rd}$ order MEFSS for example, there is a 270° phase variation between 9 and 11GHz. The technique used to obtain the desired phase delay as a function of frequency, $\phi(f)$, is to detune the center frequency of the MEFSS by varying the dimensions $g_1$, $g_2$, and $w$ (Fig. 1). The dielectric thickness $h$ is kept constant. The detuned $f$, however, should remain within the passing band to ensure high transmission. At the same time, it is desired here that $\phi$ varies linearly with $f$. These two requirements limit the phase variation at any frequency (within the passing band) to 270°. For a planar lens simple geometry gives the aperture limitation for a given frequency and focal length as:

$$A = 2\left[\left(\frac{\Delta\varphi_{max}}{k}+l\right)^2 - l^2\right]^{\frac{1}{2}} \tag{5}$$

where $A$ is the aperture diameter, $l$ is the focal length, $k$ is the frequency dependent wave number, and $\Delta\varphi_{max}$ is the phase variation between the optical axis and the edge of the aperture. This gives an aperture constraint of $A = 26.6 cm$ at 8GHz. What is desired is the largest phase variation possible so that a large amount of control over the focus is maintained. There is, however, a tradeoff. The phase response required for the m-th zone is determined by the radial distance $r_m$ from the optical axis to a point half way between the inner and outer radius of the zone. Increasing the number of zones decreases the width of each zone due to the limitation of the aperture which means that for the outermost zone, $r_m$ approaches $\frac{A}{2}$, and therefore the maximum phase variation, for a large number of zones. However, using more zones requires fewer cells to be used in each zone, with the result that the infinite array used in the numerical modeling of a single zone is no longer a satisfactory approximation. The use of 7 zones was chosen as a reasonable compromise between the two competing effects.

The phase delay for the center zone ($z$=0) is found from the aperture (Eq. 5) and focal lengths to be:

$$\varphi_0 = k\left(\sqrt{\left(\frac{A}{2}\right)^2 + l^2} - l\right) + \varphi' \tag{6}$$

where $\varphi'$ is a constant phase to be added to all zones of the lens.
The phase delay introduced by the m-th zone, located at a radial distance $r_m$ from the optical axis, is:

$$\varphi_m = k\left(\sqrt{\left(\frac{A}{2}\right)^2 + l^2} - l\right) + \varphi' - k\left(\sqrt{r_m^2 + l^2} - l\right) \tag{7}$$

Therefore the phase difference between zones 0 and 6 is:
$$\Delta\varphi = \varphi_0 - \varphi_6 = k\left(\sqrt{r_6^2 + l^2} - l\right) \tag{8}$$
For $f$=8GHz and $l$=30cm this phase difference (which will be one of the goals of our optimization) evaluates to:
$$\Delta\varphi = 218° \tag{9}$$
For $f$=12GHz and $l$=126cm it evaluates to:
$$\Delta\varphi = 81° \tag{10}$$

It is evident from the goals set in equations (9)-(10) that the desired effect is to have phase responses that *converge* towards higher frequencies. Converging phase responses would mean that for high $f$ the phase delay differences between zones become small, and so since all zones would be phase-shifting the wave by nearly the same amount an incident plane wave would remain approximately planar, i.e. that it's focal length is very large which is the behavior we want at high $f$.

To ensure a high degree of accuracy in finding parameters that not only match the desired phase at 8 and 12GHz but also have a linear phase response a computerized optimization was carried out in two steps. First, a 3D parameter scan was conducted: the dimensions $w$, $g_1$, and $g_2$ were varied and the corresponding phase shift $\varphi$ was calculated for various frequencies. A contour plot of $\varphi$ in 3D (as a function of $w$, $g_1$, and $g_2$) is shown in Fig. 4 for $f=10$GHz. Similar sets of data were generated for $f=8, 8.5, 9,…,12$GHz. For each zone then, Mathematica can be used to locate the set of parameters that approximately minimizes the goal function value (GFV) $G_i$:

$$G_i = \sum_{f=8,8.5,...12\text{GHz}} [\varphi_f(w, g_1, g_2) - \varphi_f^{(i)}]^2 \qquad (11)$$

where $\varphi_f(w, g_1, g_2)$ is the *actual* phase value obtained in CST simulations at frequency $f$ for a specific set of parameters $w$, $g_1$, and $g_2$. By contrast, $\varphi_f^{(i)}$ is the *desired* phase value at frequency $f$ and for zone $i$.

Minimizing $G_i$ is equivalent to a least-square minimization of the discrepancy between the actual and desired phases at frequencies $f=8,8.5,…12$GHz in the X-band. Note however, that Eq. 6-8 only determine the phase differences between zones at certain frequencies and does not explicitly determine the desired phase values, $\varphi_f^{(i)}$, of an individual zone. Only in choosing the desired phase values for one zone are they then determined for the remaining zones. The desired phase response of a zone can be characterized by the phase value at 8GHz (offset) and the slope given by the difference in phase between 8-12GHz. For zone i then, using Mathematica and Eq. 11, the GFV was minimized not only by varying the parameters $w$, $g_1$, and $g_2$, but also by varying the offset and slope which changes the desired phase values $\varphi_f^{(i)}$. An optimum offset and slope was found that simultaneously minimized the GFV for zone 0 and 6. This ensured that the largest GFV returned by any of the 7 zones would be as small as possible. Using this offset and slope, the desired phase responses were evaluated for all the zones. With the desired phase responses determined, a set of parameters was found for each zone that minimized the GFV.

Once the approximate parameter minimum was found in Mathematica, further finer scans of $w$, $g_1$, and $g_2$ were performed by means of dedicated full-wave CST simulations. This led to identifying the best dimensions $w$, $g_1$, and $g_2$ for the closest match of $\varphi(f)$ with its goal for zone i. The procedure was repeated for all zones. The results, summarized in Fig. 5 and Table 1 confirm that it is possible to engineer a zoned lens of reverse chromatic aberration as desired. The results vary in their tolerance. Zone 6 for example, exhibits behavior very close to the desired phase response over the 8-12GHz frequency range except near 8GHz where there is a deviation of approximately 30°. This maximum deviation diminishes for the lower zones, but the overall behavior also becomes less linear and the phase response is seen to be alternatively above and below the desired linear response.

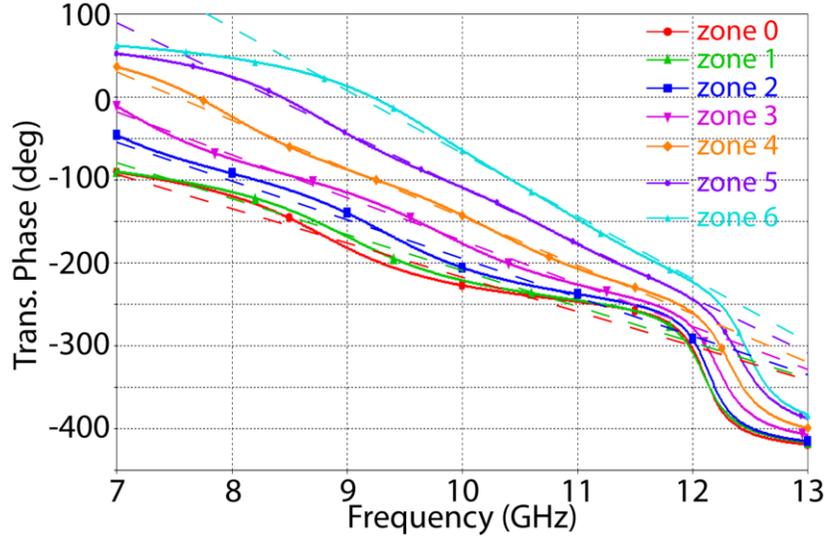

Fig. 5. Phase response of the numerically optimized zones (solid lines) and their corresponding desired phase responses (dashed lines)

Three parameters were varied for each zone: $w$, $g_1$, and $g_2$, but it is expected that a higher degree of control of $\varphi(f)$ can be achieved by adding other free parameters such as $D$, $h$, or by allowing different layers to have different dimensions, or by considering MEFSS of higher order than N=3.

**Table 1. Optimal Parameter Dimensions by Zone**

| Zone | $w$ (mm) | $g_1$ (mm) | $g_2$ (mm) |
|---|---|---|---|
| 0 | 0.245 | 0.233 | 0.540 |
| 1 | 0.199 | 0.274 | 0.510 |
| 2 | 0.763 | 0.278 | 0.511 |
| 3 | 0.861 | 0.364 | 0.498 |
| 4 | 1.193 | 0.436 | 0.500 |
| 5 | 1.485 | 0.502 | 0.494 |
| 6 | 1.931 | 0.507 | 0.514 |

## 3. Conclusion

In conclusion it has been shown by numerical scans of the phase shift introduced by a 3[rd] order miniaturized element frequency selective surface (MEFSS) as a function of its geometrical properties, that the phase response in the various zones composing a composite metamaterial lens can be manipulated to craft the chromatic aberration in a desirable manner, hereby including "reverse" chromatic aberration. Modeling was performed in the 8-12GHz range, which is of interest for millimeter wave diagnostics in low-field fusion plasmas. Future work will address the extension to higher frequencies, relevant to fusion devices operating at higher magnetic field, as well as to non-fusion applications.

**Acknowledgments**

Author WJC wishes to thank Chien Hao Liu and Meng Li for their help with the optimization software, Mark A. Thomas for fruitful discussions and technical assistance and John S. Sarff and Jay K. Anderson for their support. NB is supported by a Young Investigator Program (YIP) Grant from the Air Force Office of Scientific Research (Award No. FA9550-11-1-0050) and by the National Science Foundation (Award No. ECCS-1101146)